\documentclass[aps,prl,twocolumn,amsmath,amssymb,floatfix, superscriptaddress]{revtex4-2}
\usepackage{tabularx}
\usepackage{bm}
\usepackage{euscript}
\usepackage{epsfig,psfrag,subfigure}
\usepackage{graphicx}
\usepackage{color}
\usepackage{amsfonts}
\usepackage{exscale}
\usepackage{amsbsy}
\usepackage{multirow}
\usepackage[normalem]{ulem}
\usepackage[colorlinks=true,linkcolor=blue,citecolor=blue, urlcolor=blue,anchorcolor=blue]{hyperref}

\begin{document}

\title{Pair-density-wave superconductivity: a microscopic model on 2D honeycomb lattice}

\author{Yi-Fan Jiang}
\email{jiangyf2@shanghaitech.edu.cn}
\affiliation{School of Physical Science and Technology, ShanghaiTech University, Shanghai 201210, China}
\author{Hong Yao}
\email{yaohong@tsinghua.edu.cn}
\affiliation{Institute for Advanced Study, Tsinghua University, Beijing 100084, China}

\begin{abstract}
Pair-density-wave (PDW) is a long-sought exotic state with oscillating superconducting order without external magnetic field.
So far it has been rare in establishing a 2D microscopic model with PDW long-range order in its ground state.
Here we propose to study PDW superconductivity in a minimal model of spinless fermions on the honeycomb lattice with nearest-neighbor (NN) and next-nearest-neighbor (NNN) interaction $V_1$ and $V_2$, respectively.
By performing a state-of-the-art density-matrix renormalization group (DMRG) study of this $t$-$V_1$-$V_2$ model at finite doping on six-leg and eight-leg honeycomb cylinders, we showed that the ground state exhibits PDW ordering (namely quasi-long-range order with a divergent PDW susceptibility).
Remarkably this PDW state persists on the wider cylinder with 2D-like Fermi surfaces (FS).
To the best of our knowledge, this is probably the first controlled numerical evidence of PDW in systems with 2D-like FS.
\end{abstract}
\date{\today}

\maketitle

In conventional BCS theory, two electrons pair with zero center-of-mass momentum and pairing amplitude respects translational symmetry \cite{BCS}. Nonetheless, it was proposed later by Fulde-Ferrell-Larkin-Ovchinnikov (FFLO) that superconducting states with finite-momentum pairing can occur in weakly interacting systems with external magnetic field \cite{Larkin1965, Fulde1964}.
Although evidences of the FFLO state has been reported in a few quantum materials \cite{Gloos1993, Modler1996, Yamashita1997, Radovan2003, Bianchi2003, Cho2017, Agosta2018, Kasahara2020}, it remains elusive to unambiguously establish FFLO states experimentally.

Recently pair-density-wave (PDW) was introduced as an exotic superconducting state with finite-momentum pairing (namely its pairing amplitude oscillates in real space), without the need of applying external magnetic field \cite{Agterberg2020}.
Increasing interest has been focused on evidences of PDW in quantum materials such as underdoped cuprate superconductors \cite{Li2007,Berg2007, Agterberg2008, Berg2009np,Fradkin2015,Hamidian2016, Ruan2018, Edkins2019, Du2020, Li2021}, iron-based superconductors \cite{Zhao2023,Liu2023}, heavy-fermion materials \cite{Gu2023,Aishwarya2023}, and kagome superconductors \cite{Chen2021}.
These novel PDW states have spatial pairing modulation similar to the FFLO state, but appear without external magnetic field.
It is widely believed that PDW in those systems emerges mainly due to strong correlations. Nonetheless, it remains a challenge to obtain controlled or reliable evidences of PDW in two and higher dimensional microscopic models, despite mean field and other studies have shown that PDW states may be realized in ground states of microscopic models \cite{Himeda2002, Raczkowski2007, Wu2007, Aperis2008, Yang2009, Roy2010, Loder2011, You2012, Cho2012, Lee2014, Soto2014, Soto2015, Jian2015, Wang2015, Wardh2017, Wardh2018, Han2020, Tommy2020,chakraborty2021odd, setty2022, Jin2022, Han2022, Coleman2022,Tommy2022, Shaffer2023, Shaffer2023triplet, Wu2023Raghu, Wu2023Yao, JiangBarlas2023, Setty2023, schwemmer2023pair}. 

So far, controlled numerical evidence of PDW has been shown in density-matrix renormalization group (DMRG) studies of quasi-1D models, including Refs. \cite{Berg2010, Jaefari2012, Venderley2019,Zhang2022, Zhou2022, chen2023singlet, Jiang2023}. 
(Note that DMRG studies of doped candidate quantum spin liquid models have observed quasi-long-range PDW correlation with finite PDW susceptibility \cite{Xu2019,Peng2021,Peng2021hub}.)
More recently, DMRG evidence of PDW is observed in the strong coupling limit of the Holstein-Hubbard model with a single 1D-band crossing on Fermi surface (FS) \cite{Huang2022}. Evidence of PDW in 2D models with 2D-like FS has not been established yet in controlled DMRG calculations.

\begin{figure}[bt]
\centering
\includegraphics[width=0.85\linewidth]{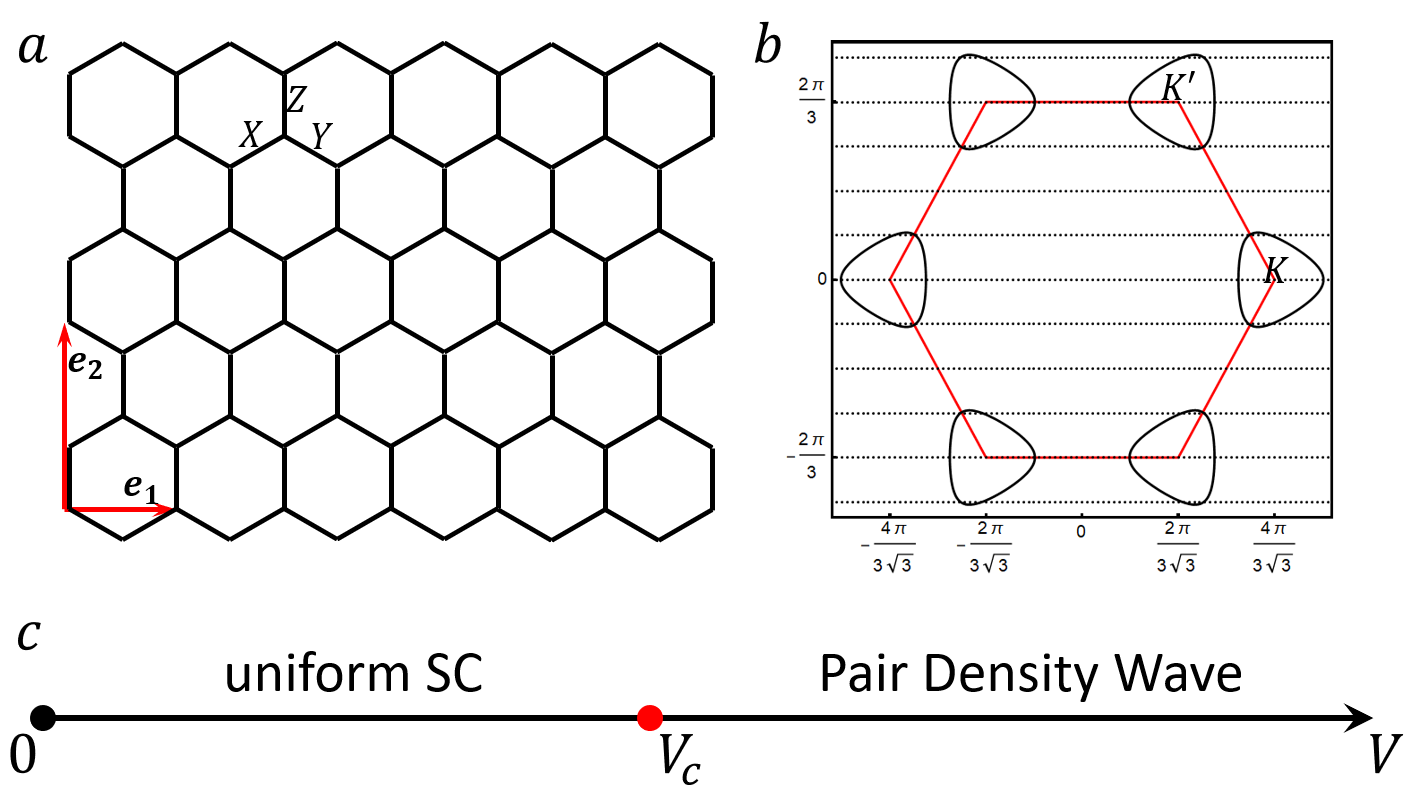}
\caption{(a) The example of honeycomb lattice with $N=2\times 6 \times 6$. The boundary condition is open along $\hat{e}_1$ direction but periodic along $\hat{e}_2$. $X$, $Y$ and $Z$ define three types of bonds. (b) The Fermi surface of the doped spinless model: it has two hole pockets around the $K$ and $K'$ points. Dash lines are the eight cuts with different $k_y$ momentum for eight-leg cylinder. (c) The quantum phase diagram of the spinless $t$-$V_1$-$V_2$ model with light doping $\delta$, as a function of interaction strength $V$. Here we mainly focus on $\delta=1/9$, $V_1=-V$, and $V_2=V/2$. 
Our study further 
demonstrates that the PDW phase can
emerge in a finite range of $V_2/V_1$ ratios and doping levels  
(see the SM for more details). }
\label{Fig:phasediagram}
\end{figure}

Here we employ DMRG \cite{White1992} to explore possible evidences of PDW in a 2D fermion model on honeycomb lattice with attractive nearest-neighbor (NN) $V_1$ and repulsive next-nearest-neighbor (NNN) interaction $V_2$ away from half filling. 
It was previously suggested in Ref. \cite{Jian2015} that, at half filling, PDW with intra-valley pairing can emerge in the ground state of the $t$-$V_1$-$V_2$ model; moreover the PDW transition features emergent supersymmetry. 
It was argued that, {\it at half filling}, PDW wins over uniform pairing because PDW can fully gap out the Dirac cones but uniform BCS pairing cannot.
Nonetheless, it remains elusive if at half filling PDW emerges from exact-diagonalization study of the half-filling model on a small cluster 
\cite{Capponi2015,Capponi2017}. 
So, it is natural to ask whether PDW occurs in the $t$-$V_1$-$V_2$ model from large-scale calculations on systems with much larger size approaching 2D. 

From state-of-the-art DMRG study of the $t$-$V_1$-$V_2$ model {\it away from half filling}, we shows reliable evidences of PDW in the doped model with two hole pocket Fermi surfaces around the two Dirac points $\pm K$.   
Specifically, we find pairing between two electrons on the same hole pocket (namely intra-valley pairing) emerges in the ground state, leading to a PDW state. 
Our DMRG calculations, greatly enhanced by GPU accelerations, enable us to simulate the model on the eight-leg cylinders featuring 2D-like FS for the first time; our simulations on the wider cylinders remarkably revealed that the PDW state persists in the model with a 2D-like FS, which is beyond 1D systems. 
To the best of our knowledge, this is probably the first controlled numerical evidence substantiating the existence of PDW in systems exhibiting a 2D-like FS.
Moreover, a 2D $t$-$V_1$-$V_2$ spinless fermion model hosting a PDW ground state may be potentially realized in various spin-polarized electron systems such as twisted moire systems \cite{Cao2018a, Cao2018b, Liu2020, Slagle2020, Devakul2021}, ultracold fermions \cite{Hulet2006, Tilman2014}, and Fe-based compounds with strong Hund's coupling \cite{Kee2017}.

{\bf Model:}
We now consider the spinless fermion model on honeycomb lattice with density-density interactions described by the following Hamiltonian
\begin{eqnarray}\label{Eq:Ham}
H =-t\sum_{\left<ij\right>} (c^\dagger_i c_j + h.c.) + V_1\sum_{\left<ij\right>} n_i n_j
 + V_2\sum_{\left<\left<ij\right>\right>} n_i n_j, ~~
\end{eqnarray}
where $c^\dagger_{i}$ is the electron creation operator on site $i=(x_i,y_i)$ and $n_{i}=c^\dagger_{i}c_{i}$ is electron number operators.
Here $t$ denotes NN hopping amplitude and we set $t=1$ as the energy unit.
The strength of the NN density-density interaction is labeled by $V_1$ and the NNN interaction $V_2$. 
We focus on the case of $V_1=-V$ and $V_2=\frac{V}{2}$, and look for possible PDW ordering by varying $V$. The PDW order in the models with  other $V_2/V_1$ ratios and different doping concentrations is also studied, as detailed in Supplemental Materials (SM).

We employ DMRG to study the $t$-$V_1$-$V_2$ model on honeycomb lattice with cylindrical geometry, as depicted in Fig. \ref{Fig:phasediagram}(a), where we take the periodic boundary condition in the $\mathbf{e}_2=(0,\frac{3}{2})$ direction and open boundary condition in the $\mathbf{e}_1=(\sqrt{3},0)$ direction. 
Here, we focus on cylinders with width $L_y$ and length $L_x$, where $L_y$ and $L_x$ are number of unit cells along the armchair chain in the $\mathbf{e}_2$ and zigzag chain in the $\mathbf{e}_1$ directions, respectively. 
There are $N_c=L_xL_y$ number of unit cells (and $N=2L_xL_y$ number of lattice sites). 
Suppose $N_e$ denotes the number of electrons; at half filling, $N_e=N_c$ which is $N/2$. 
The concentration of doped holes is defined as $\delta=\frac{N_h}{N_c}$ with $N_h=N_c-N_e$. 
In the present study, we focus on $L_y=6$ and $L_y=8$ cylinders with $L_x$ up to 48 at hole doping concentration $\delta\leq 1/8$. 
We perform up to 100 sweeps and keep up to $33,000$ U(1) DMRG block states to obtain a typical truncation error $\epsilon \lesssim 5\times 10^{-6}$. 
Further details of the numerical simulations are provided in the SM.

\begin{figure*}[bt]
\centering
\includegraphics[width=0.95\linewidth]{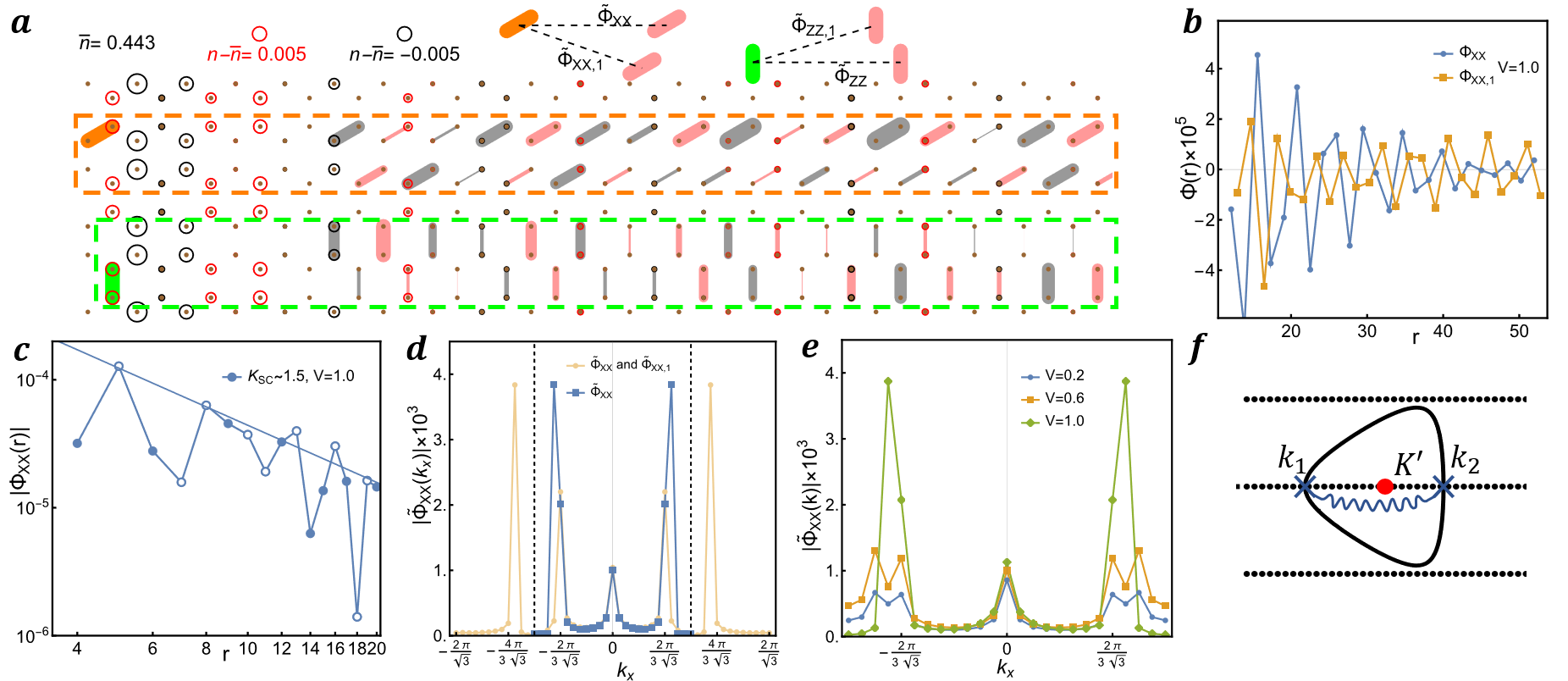}
\caption{DMRG results of PDW on the 6-leg cylinder: (a) Charge density profile and scaled pair-pair correlation functions $\tilde{\Phi}(r)$ of $V=1.0$ model on $2\times 48 \times 6$ cylinder with doping concentration $\delta\sim 11.1\%$. Radius of circles show the difference $n(x,y)-\bar{n}$ where $\bar{n}$ is the average density, and thickness of bonds represent the amplitude of four types of $\Phi(r)$ illustrated on the top of figure. Negative values are marked black. (b) Pair-pair correlation function $\Phi_{\text{XX}}(r)$ and $\Phi_{\text{XX,1}}(r)$ of the same model. (c) Long-range behavior of $\Phi_{\text{XX}}(r)$ fitted by the power-law function $\sim r^{-K_{sc}}$, empty circles denote negative values. 
(d) Fourier transform of scaled correlation functions: Orange line presents the results based on $\tilde{\Phi}_{\text{XX}}(r)$ and $\tilde{\Phi}_{\text{XX,1}}(r)$ and blue line shows the results obtained from $\tilde{\Phi}_{\text{XX}}(r)$ alone. Dashed lines illustrate the folded Brillouin zone. 
(e) Fourier transform of scaled correlation $\tilde{\Phi}(r)=\Phi(r)/r^{-K_{sc}}$ for the models with $V=0.2 \sim 1.0$ in folded Brillouin zone. 
(f) Intra-valley pairing of fermions within the same hole pocket.
}
  \label{Fig:ly=6result}
\end{figure*}

{\bf Quantum phase diagram: }
At doping slightly away from half filling, the Fermi surface of the model in 2D becomes two hole pockets around two Dirac points $\pm K$ as shown in Fig.~\ref{Fig:phasediagram}(b). 
In the non-interacting limit ($V=0$), the single-particle dispersion is denoted as $\epsilon(\vec k)=-2t(\cos k_x+\cos k_y)-\mu$, where $\mu$ denotes the chemical potential; the susceptibility of uniform SC diverges logarithmically with temperature due to the nesting of the Fermi surface in the {\it particle-particle} channel, namely $\epsilon(\vec k)=\epsilon(-\vec k)$, guaranteed by the inversion or time-reversal symmetry of the model.
Consequently, when the interaction $V$ is finite but sufficiently weak, uniform SC with usual zero-momentum pairing is expected in the ground state.
Here, we are more interested in the regime beyond the weak coupling physics.  
Indeed, when $V$ exceeds a critical value, our DMRG simulations show that the pair-pair correlation function starts to exhibit intense sign-changing oscillation at long distance, i.e. forming a PDW state. 
Correspondingly, the Fourier transformation of the scaled pair-pair correlation function exhibits two sharp peaks at non-zero momentum $Q\sim \pm 2K$, which implies that Cooper pairs in the PDW state consist of fermions from the same hole pocket.
Interestingly, this PDW phase can be solely induced by $V_1$ and extends to a range of doping level around $10\%$. 
We would like to emphasize that the PDW here is incommensurate away half filling, importantly distinct from the possible commensurate PDW realized at half filling. 

Here we would like to emphasize the key importance of 8-leg cylinder. As the cylinders respects translation symmetry along $y$, $k_y=2\pi\frac{n_y}{L_y}$ with integer $n_y \in (-\frac{L_y}{2},\frac{L_y}{2}]$ is a good quantum number; $k_y$ can be employed to label bands and the total number of $k_y$ bands is $L_y$.  
As width increases, the small Fermi pockets could be crossed by $k_y$ bands with both $k_y=0$ and non-zero $k_y$ such that the system starts to approach to the 2D limit when more and more $k_y$ bands cross the Fermi level.
This condition is satisfied on the 8-leg systems which is studied in the present work. 
As a consequence, we start to observe the 2D-like behavior in such system, e.g. the density profile shows a tendency to restore the $C_6$ rotation symmetry of the 2D lattice on the eight-leg cylinders although the 8-leg cylinder still slightly breaks the $C_6$ symmetry.
Remarkably, our DMRG study shows that the spatial oscillation of the pair-pair correlation function is still robust on 8-leg cylinders with moderate interaction and there are the pronounced peaks around $\pm 2K$ points.
These numerical evidences indicate that the PDW ordering could persist in the minimal model Eq. (\ref{Eq:Ham}) even when it starts to approach 2D with 2D-like FS. 
The quantum phase diagram of the doped 2D spinless model obtained from our DMRG study is sketched in Fig. \ref{Fig:phasediagram}(c).

{\bf PDW on six-leg cylinders: }
Possible SC ordering in the $t$-$V_1$-$V_2$ model in Eq. (\ref{Eq:Ham}) can be characterized by equal-time pair-pair correlations defined as
\begin{eqnarray}
    \Phi_{\alpha \beta}(r)=\left<\hat{\Delta}_{\alpha}^{\dagger}(x_0,y_0) \hat{\Delta}_{\beta}(x_0+r,y_0)\right>,
\end{eqnarray}
where $\hat{\Delta}_{\alpha}^{\dagger}(x,y) = c^{\dagger}_{(x,y)} c^{\dagger}_{(x,y)+\alpha} $ is a pair creation operator on bond $\alpha=X$, $Y$, $Z$, as illustrated in Fig. \ref{Fig:phasediagram}(a). 
Here $(x_0, y_0)$ is the reference bond taken as $x_0 \sim 5$ unit cells away from the open boundary and $r$ is the distance between two bonds in the $\mathbf{e}_1$ direction. 
The colored bonds in Fig. \ref{Fig:ly=6result}(a) depict four different types of pair-pair correlations between bonds on same zigzag chain ($\Phi_{\text{XX}}$, $\Phi_{\text{ZZ}}$) and adjacent chains ($\Phi_{\text{XX,1}}$, $\Phi_{\text{ZZ,1}}$). At long distance, all four correlations exhibit sign-changing oscillations that alternate between negative (black) and positive (red) bonds. A direct comparison of correlations $\Phi_{\text{XX}}(r)$ and $\Phi_{\text{XX,1}}(r)$ is presented in Fig. \ref{Fig:ly=6result}(b), where $r$ is the distance between two bonds along the $\mathbf{e}_1$ direction. Both correlation functions possess similar wavelength of $\sim 3$ unit cells and exhibit decaying behaviors that can be fitted by power-law functions. However, there is a roughly $\pi$-phase shift between the oscillations of $\Phi_{\text{XX}(r)}$ and $\Phi_{\text{XX,1}(r)}$. Combined these key features together, the pair correlation $\Phi(r)$ can be described generally as: 
\begin{eqnarray}
  \Phi(r) \sim r^{-K_{sc}} \cos (Q r+ \theta),
  \label{Eq:pdw}
\end{eqnarray}
where $Q$ is a non-zero ordering vector and $K_{sc}$ is the power-law decay exponent. For $V=1.0$, the extracted exponent $K_{sc}=1.5(2)$ from Fig. \ref{Fig:ly=6result}(b), which suggests that the corresponding SC susceptibility $\chi_{sc} \sim T^{- (2 - K_{sc}) }$ diverges as the temperature $T\rightarrow 0$, under the assumption of emergent Lorentz symmetry in low energy effective theories of 1+1D systems \cite{Arrigoni2004}.
This establishes that the lightly doped spinless fermion model on six-leg cylinders has quasi-long-range SC correlations in the relatively strong $V$ region. 

Using the extracted exponent $K_{sc}$, we further define the scaled SC correlation function $\tilde{\Phi}(r)=\Phi(r)/r^{-K_{sc}}$ to investigate the spatial oscillation of the SC ordering.
To determine the ordering vector $Q$ in Eq. \ref{Eq:pdw}, we calculate the Fourier transform $\tilde{\Phi}(k)=\frac{1}{N}\sum_{r} e^{-ikr}\tilde{\Phi}(r)$ of the combination of $\tilde{\Phi}_{\text{XX}}(r)$ and $\tilde{\Phi}_{\text{XX,1}}(r)$ shown in Fig. \ref{Fig:ly=6result}(a) and (b). The ordering vector $Q \sim K=\{\frac{4\pi}{3\sqrt{3}}, 0\}$ is extracted from the pronounced peak of the orange line in Fig. \ref{Fig:ly=6result}(d). 
The pair momentum $Q \sim K$ can be explained by the intra-valley pairing of two fermions from the same hole pocket at $\pm K$ point. Note that the observed PDW is incommensurate as $Q$ does not exactly match the $\pm K$ point. This deviation can arise from the asymmetric Fermi points illustrated in Fig. \ref{Fig:ly=6result}(f): the momentum of the Cooper pair is $k_1+k_2 < 2 K'$, which is equivalent to $k_1+k_2+3K < 2 K' + 3 K = K$ as expected. It is intriguing that this incommensurate feature remains in the strong coupling region. Additionally, there are two subleading peaks around $\pm K/2$ points in $\tilde{\Phi}(k)$, which reflect the amplitude discrepancy between $\Phi_{\text{XX}}$ and $\Phi_{\text{XX,1}}$. 
The subleading peak can be removed by considering the Fourier transform of $\tilde{\Phi}_{\text{XX}}(r)$, where a single pair of prominent peaks corresponding to the momentum $Q$ appears in the folded Brillouin zone (BZ). 

Once the leading ordering vector $Q$ is determined, we investigate the evolution of the PDW ordering as $V$ and $L_x$ varied. 
We focus on the peaks of $\tilde{\Phi}_{\text{XX}}(k)$ in the folded BZ. The $\tilde{\Phi}_{\text{XX}}(k_x)$ of the systems with interaction $V=0.2, 0.6, 1.0$ are summarized in Fig. \ref{Fig:ly=6result}(e). 
The single peak at $k=0$ corresponds to the uniform BCS pairing which is dominant in the weak $V$ region.
When the interaction is gradually increased from $V=0.2$ to $V=1.0$, we observe that two peaks at finite-momentum  $Q\sim\pm 2K$ emerge, and they become more dominant than the $k=0$ one after the interaction exceeds a critical value $V_c \sim 0.6$, indicating the onset of the PDW state.
The existence of PDW phase is further supported by the finite-size scaling of the peaks in the $V=1.0$ model. As $L_x$ increases, the finite-momentum peaks of $\tilde{\Phi}(k)$ become more pronounced, while the peak at zero momentum is suppressed (see SM for details), which suggests a dominating PDW order in the strong interaction regime.

We further investigate the potential PDW phase in systems with other $V_2/V_1$ and doping concentrations. Interestingly, our calculations show that for sufficiently large $V_1$, the PDW phase could persist on the six-leg cylinder even when $V_2=0$. For instance, for the $V_2=0$ model on $L_x=24$ cylinder with $\delta=8.33\%$, a dominant peak of $\tilde{\Phi}(k)$ arises when $V_1$ is larger than $\sim 1.0$ (see SM for details). The momentum of the peaks does not obviously depend on the doping concentration, agreed with the picture of intra-valley pairing depicted in Fig. \ref{Fig:ly=6result}(f). 


The charge density of the ground-state is $n(x,y)=\langle \hat{n}(x,y)\rangle$. For $V=1.0$,  $n(x,y)$ on the left part of the $L_x=48$ cylinder is shown in Fig. \ref{Fig:ly=6result}(a).
The density profile respects translation symmetry along $\mathbf{e}_2$ direction but has spatial oscillation with period $\lambda_c \sim 3$ along the $\mathbf{e}_1$ direction which decays quickly in the bulk. 
This spatial decay can be interpreted as the Friedel oscillation induced by the open boundary of the cylinder: $n(x)=A \cos(Q_c x +\phi) x^{-K_c/2} + \bar n$, where $Q_c=2\pi/\lambda_c$ is the CDW ordering vector and $K_c$ is the Luttinger exponent characterizing the density correlation \cite{White2002}. 
We obtain $K_c\sim 1.7(3)$ which indicates a subdominant CDW correlation in the PDW phase (see SM for details).

\begin{figure}[bt]
\centering
\includegraphics[width=\linewidth]{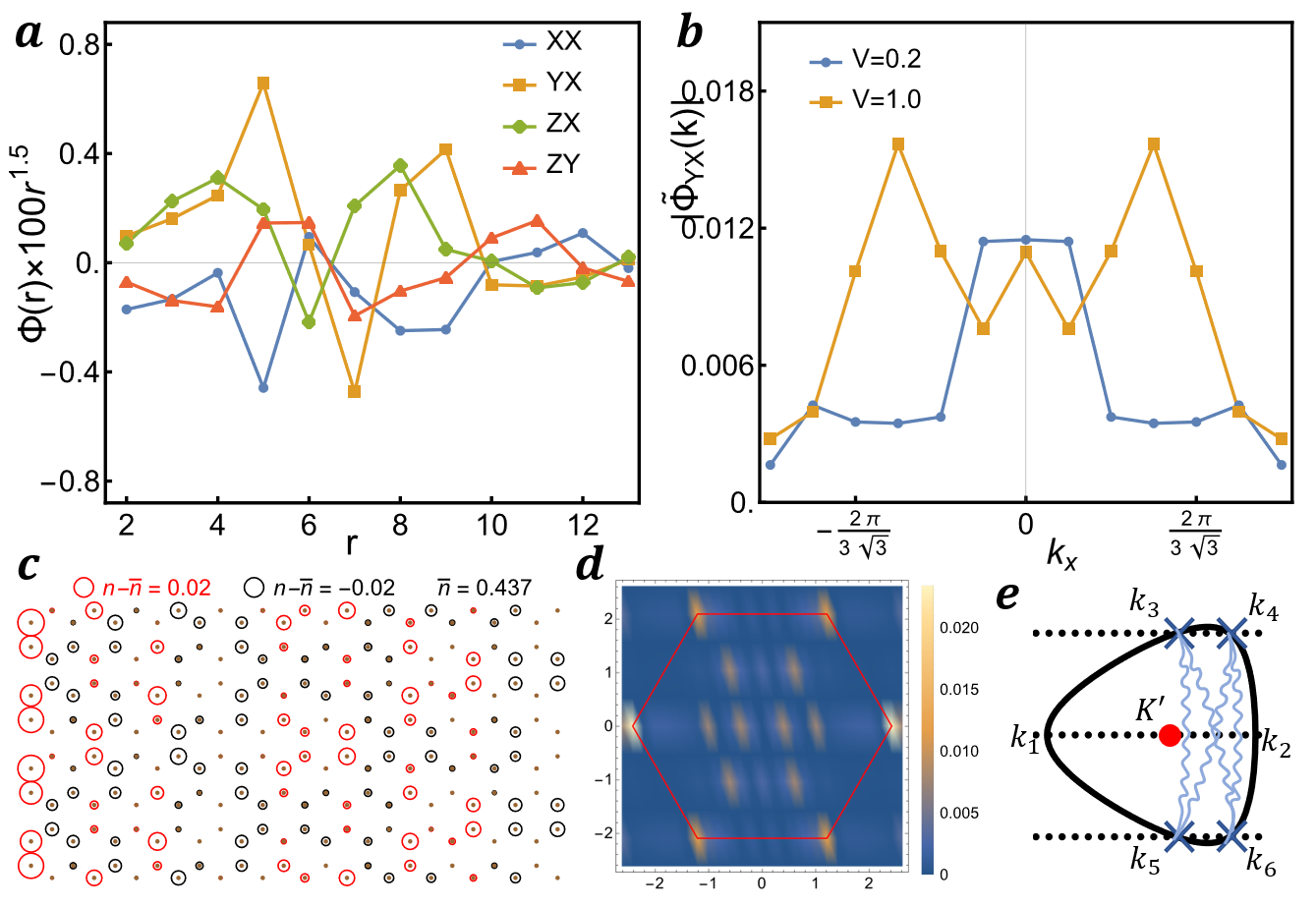}
\caption{PDW on 8-leg cylinders: (a) Scaled pair-pair correlation function $\tilde{\Phi}(r)=\Phi(r)/r^{-1.5}$ for the model with $V=1.0$, $L_x=21$, and doping $\delta\sim 12\%$. (b) Fourier transform of $\tilde{\Phi}(r)$ for the models with $V=0.2$ and $1.0$. (c) Charge density profile of the model with $V=1.0$. Radius of circles show the difference $n(x,y)-\bar{n}$ where $\bar{n}$ is the average density, and black circle means negative value. (d) Fourier transform of the charge density profile in (c). (e) Examples of intra-valley pairing between multiple Fermi points for $L_y=8$.}
\label{Fig:ly=8result}
\end{figure}

{\bf PDW on wider systems: }
In many previous studies, evidences of PDW are observed mainly in models on 1D chains and narrow cylinders \cite{Berg2010,Jaefari2012, Venderley2019, Xu2019,Peng2021,Peng2021hub,Jiang2023}.
Our study on six-leg cylinders is among the widest, but it is still essentially 1D in the sense that only the band with $k_y=0$ crosses the 2D-limit Fermi surface. 
To have more bands crossing the 2D-limit Fermi surface, one needs to go to wider cylinders. 
So, we further studied the lightly doped model on eight-leg cylinders for which multiple bands (both $k_y=0$ and $k_y\neq 0$) can cross the hole pocket Fermi surface, as shown in Fig. \ref{Fig:phasediagram}(b), such that it looks like a 2D FS.
As a direct reflection of this dramatic change in low-energy physics, our DMRG simulation on the 8-leg model starts to observe non-trivial physical properties beyond 1D.
For instance, from the density profile $n(x,y)$ of $V=1.0$ and $\delta=12\%$ model shown in Fig. \ref{Fig:ly=8result}(c), we can see that the translation symmetries along both $\mathbf{e}_1$ and $\mathbf{e}_2$ are broken, which is distinct from that of 6-leg cylinder.
The Fourier transform of the charge density $n(\bold k)=\frac{1}{N} \sum e^{i \bold{k \cdot r}} n(\bold{r})$
in Fig. \ref{Fig:ly=8result}(d) reveals details of additional symmetry breaking along $\mathbf{e}_2$ for 8-leg case: besides the two leading peaks at $\pm K$ points there are four additional sharp peaks appear at $C_3$-rotation equivalent points of $\pm K$, indicating that the density profile tends to restore the $C_6$ rotation symmetry of the 2D lattice.

After observing the qualitative change in charge density profile for the wider system, one may naturally ask whether the PDW ordering persists when we extend the narrow cylinders to the 2D-like system.
To answer this question, we investigate the possible PDW on 8-leg cylinders and positive evidences of robust PDW are obtained in our DMRG calculation of the 8-leg system.
Due to the enormous amount of low-energy states introduced by multiple Fermi points, getting converged results on 8-leg cylinders is much more challenging computationally.
To improve reliability of the DMRG results, we study cylinders with length up to $L_x=21$ and push the bond dimension to $m=33000$, which is state-of-the-art. 
Fig. \ref{Fig:ly=8result}(a) shows the long-distance behavior of the scaled pair-pair correlation function $\tilde{\Phi}(r)=\Phi(r)/r^{-1.5}$ for the $V=1.0$ and $\delta = 12\%$ model on $L_x=21$ cylinder, where we can clearly see the key feature of PDW ordering, i.e. the sign-changing oscillation, in all types of pair-pair correlation functions.
After Fourier transform, the scaled correlation $\tilde{\Phi}(k)$ displays two finite-momentum peaks. Interestingly, the location of these peaks provides insight into the underlying physics beyond 1D. As depicted in Fig. \ref{Fig:ly=8result}(f), the Cooper pair composed of fermions from 2D like FS can possess momentum larger than $2K'$, e.g. $k_3+k_6>2K'$. Consequently, we observe in Fig. \ref{Fig:ly=8result}(b) that the peaks of $\tilde{\Phi}(k)$ have shifted from right side of $K/2$ (in six-leg case) to the left side. These peaks are replaced by a single peak at zero momentum in the $V=0.2$ model, indicating that the emergence of PDW for $V=1.0$ is a direct consequence of strong interactions.

{\bf Conclusion and discussions: }
We have studied the nature of SC in the lightly doped spinless fermion $t$-$V_1$-$V_2$ model with finite Fermi surfaces and found robust signature of an incommensurate PDW state on 6-leg and 8-leg cylinders. On one hand, 
the pairing momentum close to $2K$ suggests that the intra-valley pairing plays an important role in forming the PDW state. 
On the other hand, the requirement of considerably strong interaction $V>V_c\sim 0.6$ to trigger the PDW order indicates that the PDW mechanism originates from strong coupling physics. 
It will be interesting to study in the future whether this mechanism is general, namely if it can appear in other systems featuring Dirac cones, such as the doped $\pi$-flux model on the square lattice and doped models of spinful fermions.

In the present model with doping $\delta\sim 12\%$, all the cylinders narrower than 8-leg are essentially 1D systems since their low-energy physics is characterized by the two Fermi points of a single band with $k_y=0$.
To the best of our knowledge, our study on the 8-leg cylinders with multiple bands crossing the Fermi level is probably the first one that shows robust evidence of an incommensurate PDW ordering in a model with 2D-like Fermi surface. 
As the $t$-$V_1$-$V_2$ model on the honeycomb lattice may be potentially realized in systems such as twisted moire systems with spin polarization, e.g. Refs. \cite{Liu2020, Slagle2020, Devakul2021}, it could provide a promising arena to explore incommensurate PDW ground states in the future.

\emph{Acknowledgments:} We would like to thank Steve Kivelson for helpful discussions. This work is supported in part by National Key R$\&$D Program of China under Grant Nos. 2022YFA1402703 and 2021YFA1400100, the Innovation Program for Quantum Science and Technology under Grant No. 2021ZD0302502, Shanghai Pujiang Program under Grant No. 21PJ1410300, and the NSFC under Grant Nos. 12347107 and 12334003. H. Y. acknowledges the support in part by the Xplorer Prize through the New Cornerstone Science Foundation. 

\bibliography{Refs}

\clearpage

\begin{widetext}

\renewcommand{\theequation}{S\arabic{equation}}
\setcounter{equation}{0}
\renewcommand{\thefigure}{S\arabic{figure}}
\setcounter{figure}{0}
\renewcommand{\thetable}{S\arabic{table}}
\setcounter{table}{0}

\section{Supplemental Material}

\subsection{A. Numerical Details}
In the present work, we performed the GPU-accelerated DMRG simulations with charge U(1) symmetry to study the ground-state properties of the $t$-$V_1$-$V_2$ model. 
The matrix multiplication heavily used in Lanczos algorithms for solving ground-state of the effective Hamiltonian and the truncation of the Hilbert space can be greatly accelerated by GPU processors. 
To obtain reliable pair-pair correlation functions on wide cylinders, we performed an 
extrapolation for correlation functions to the zero truncation-error limit $\epsilon \rightarrow 0$ (i.e. $m\rightarrow \infty$). 
For example, on $L_x=15$ and $L_y=8$ cylinder, we first calculate correlation $\Phi_{ZZ}(\epsilon)$ by keeping $m = 22000 \sim 33000$ states for each $r$. 
For each number of states $m$, we performed at least 5 DMRG sweeps to reach converged results. 
Then, an extrapolation using a second-order polynomial $\Phi_{ZZ}(\epsilon) = \Phi_{ZZ} + a_1 \epsilon + a_2 \epsilon^2$ is applied to extract $\Phi_{ZZ}$ to the zero truncation-error limit for each $r$. Here $\epsilon$ is the truncation-error associated with the number of states $m$, and $a_1$ and $a_2$ are fitting parameters. As shown by the dense data points in Fig.~\ref{AFig:trunc_err}(a), the fitting qualities are quite good even for long-distance correlations. We further present the finite-truncation error extrapolation for correlation functions $\Phi_{YX}(r)$ and $\Phi_{ZY}(r)$ on the $L_x=21$ and $L_y=8$ cylinder in Fig. \ref{AFig:trunc_err}(b) and (c), respectively.

\begin{figure*}[bth]
\centering
\includegraphics[width=0.9\linewidth]{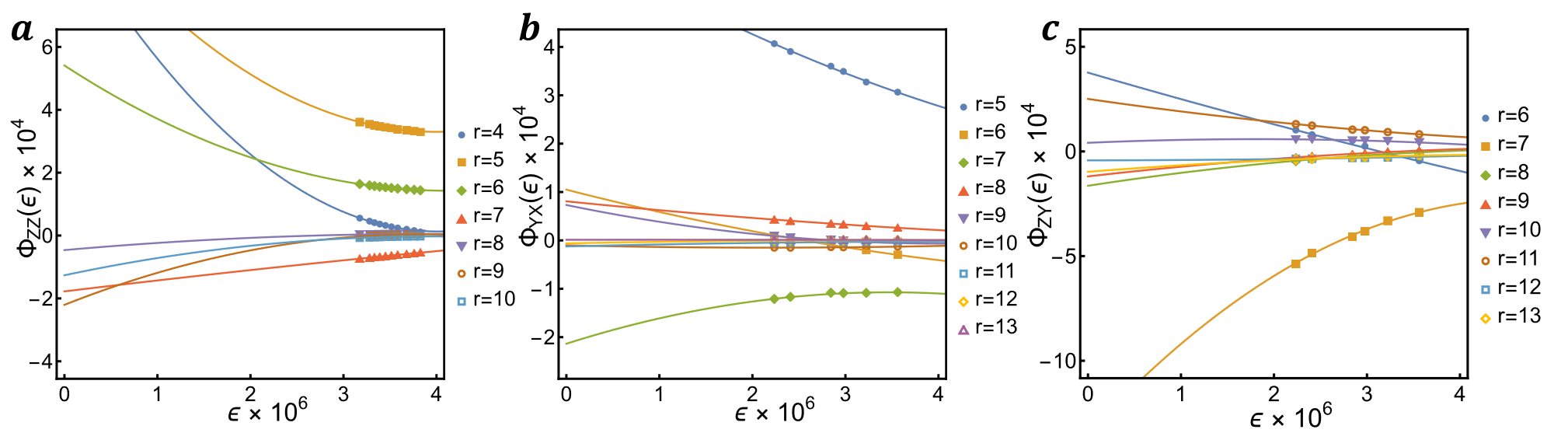}
\caption{Finite-truncation-error extrapolation for pair-pair correlation measured on 8-leg cylinders with $m=22,000 \sim 33,000$ DMRG block states. (a) The correlation $\Phi_{ZZ}(r, \epsilon)$ as a function of truncation $\epsilon$ obtained on $L_x=15$ and $L_y=8$ cylinders, $r$ is the distance between two bonds along the zigzag chain. (b) The correlation $\Phi_{ZZ}(r, \epsilon)$ on $L_x=21$ and $L_y=8$ cylinders. (c) The correlation $\Phi_{ZY}(r, \epsilon)$ on $L_x=21$ and $L_y=8$ cylinders.}
\label{AFig:trunc_err}
\end{figure*}

\subsection{B. Evidences of PDW for the case of $V_2/V_1=0$, for other cases of $V_2/V_1$, and for different doping $\delta$ }
Here we provide further numerical evidences of the PDW phase across a range of $V_2/V_1$ ratios and doping concentrations $\delta$. We choose various values of $V_2/V_1$ between $0$ and $2/3$, and perform the simulations on the $2 \times 24 \times 6$ systems with doping $\delta=1/12$ and $1/9$. One important observation is that the PDW order can be solely induced with a strong $V_1$ attraction. As shown in Fig.~\ref{AFig:otherparameter}(a), for $\delta=1/12$ and $V_2=0$, the Fourier transform of the scaled correlation function $\tilde{\Phi}_{\text{XX}}(k)$ exhibits dominant finite momentum peaks when $V_1$ exceeds a relatively larger threshold $1.0$, indicating the key role of the $V_1$ interaction in inducing PDW ordering. 
Here the same exponent $K_{sc}=1.5$ is used for all three correlation functions to ensure consistency. Similar to the finite $V_2$ case discussed in the main text, the location of the peak indicates the momentum of the Cooper pairs is around $2K$. 
The small deviation between the location of peak and $2K$ is smeared out by the relatively large interval of $k_x$ caused the small system size. 

\begin{figure*}[t]
\centering
\includegraphics[width=0.8\linewidth]{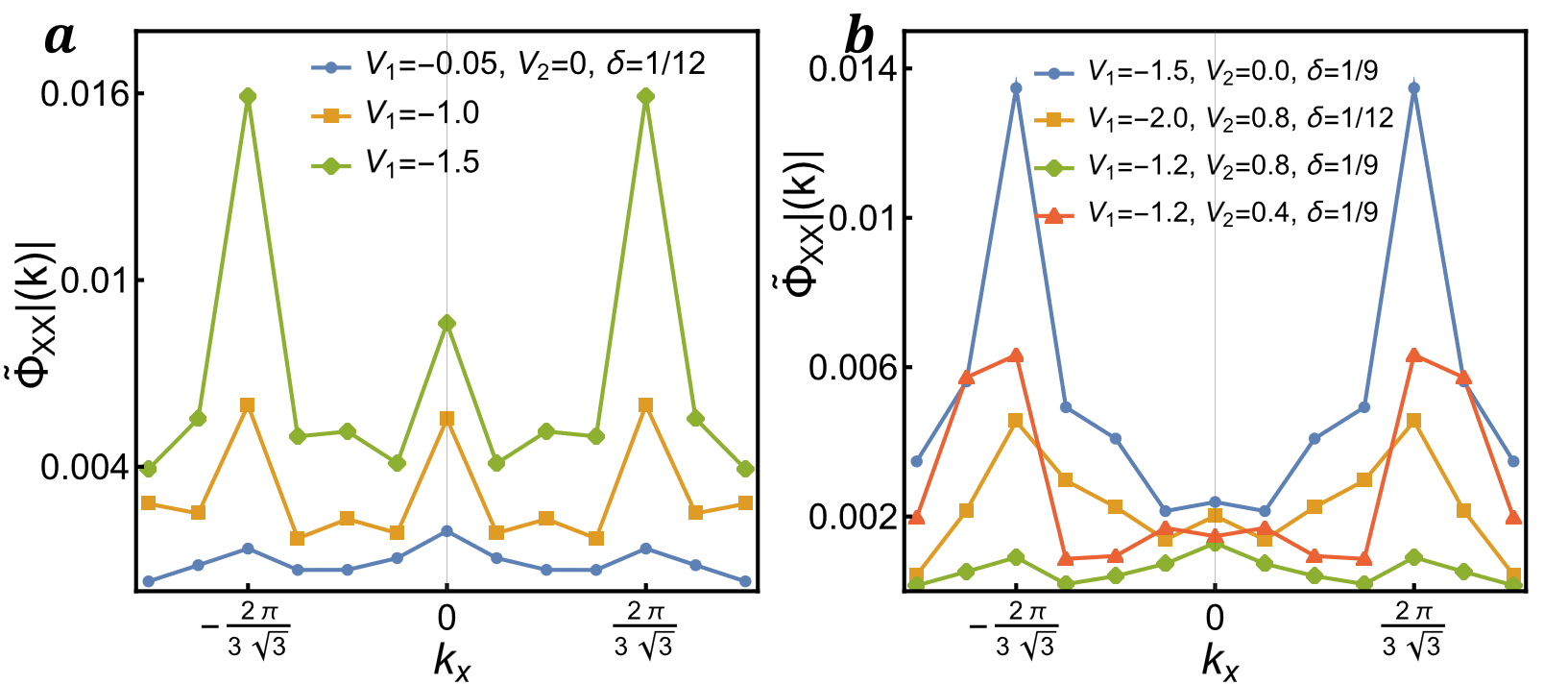}
\caption{Fourier transform of scaled correlation function for a series of $V_2/V_1$ ratios and doping $\delta$. The same exponent $K=1.5$ is used to scale the correlation $\Phi(r)$ for all the cases.}
\label{AFig:otherparameter}
\end{figure*}

We also consider other $V_2/V_1$ ratios and doping $\delta$ as summarized in Fig. \ref{AFig:otherparameter}(b). 
For  $\delta=1/9$, the PDW ordering for finite $V_1$ but $V_2=0$ is also observed, which clearly suggest that PDW can exist in a finite range of doping concentration around $10\%$. 
The effect of $V_2$ in forming the PDW state is complex. 
A weak $V_2$ is helpful to form PDW since it reduce the critical $V_1$ required for establishing the PDW order. However, there is an optimal ratio $V_2/V_1$ for PDW; when the ratio is too large, it can surpress PDW. For instance, when $V_2$ becomes comparable to $V_1$, e.g. $V_2/V_1=\frac{2}{3}$, the spatial oscillation of the pair-pair correlation is suppressed.

\subsection{C. PDW peaks in systems with different length}
To rule out possible finite size effect, we further calculate $\tilde{\Phi}(k)$ of the $V=0.2$ and $1.0$ models on the cylinder with $L_x=24 \sim 48$, as depicted in Fig. \ref{AFig:otherlength}. Again, we fix $V_1=-V$ and $V_2=V/2$, and apply same uniform scaling factor $K=1.5$ to normalize the correlation functions.
For the weakly interacting model with $V=0.2$ and $\delta=11.1\%$, we find that the finite-momentum peaks are suppressed as the length increases from $L_x=36$ to $48$; on the contrary, the zero-momentum peak gradually increases and becomes dominant in the $L_x=48$ system. This result implies that at weak coupling the uniform SC is preferred in the ground state, as expected. The opposite behavior happens for the strong interaction case of $V=1.0$, as shown in Fig. \ref{AFig:otherlength}(b); as $L_x$ increases, the finite-momentum peaks of $\tilde{\Phi}(k)$ become sharper while the zero-momentum peak is suppressed, which clearly suggests that for strong interaction $V=1.0$ the PDW with finite-momentum pairing is the only dominant SC order.

\begin{figure*}[bth]
\centering
\includegraphics[width=0.8\linewidth]{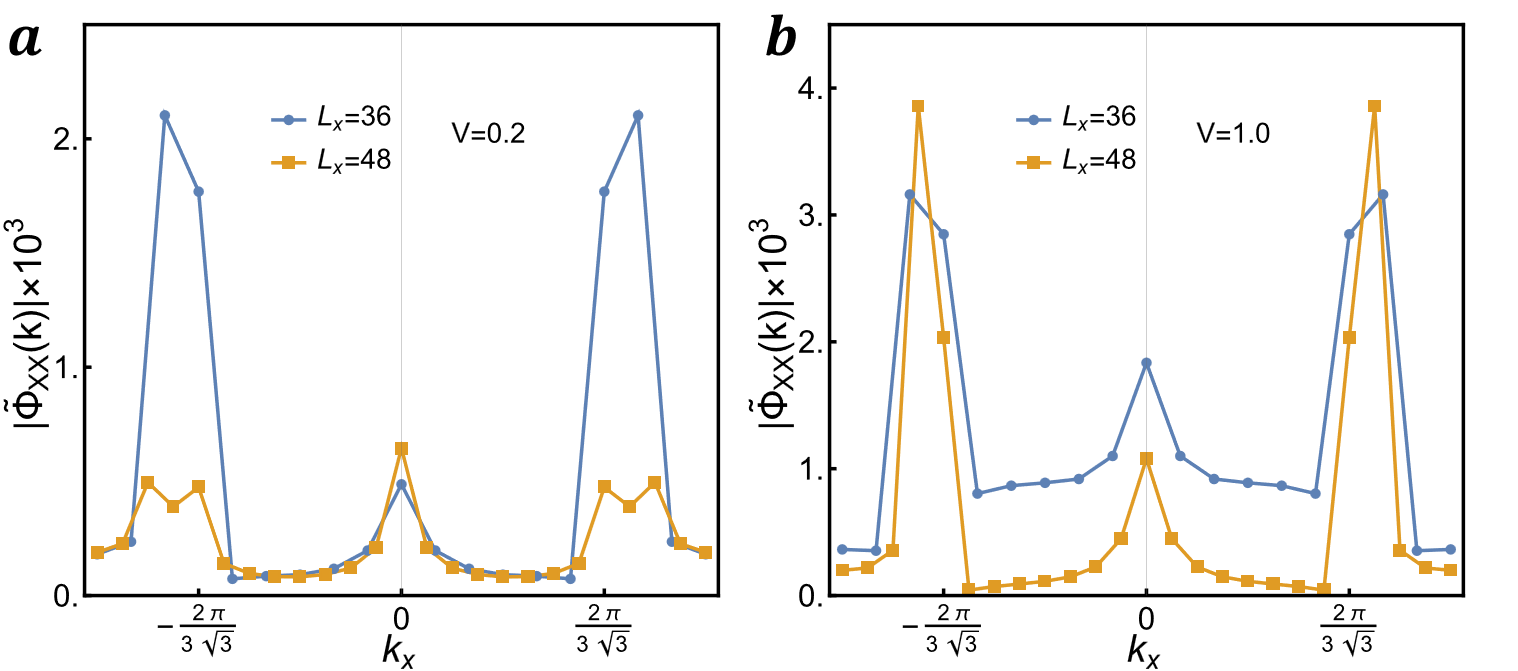}
\caption{Fourier transform of scaled correlation function for the $V=0.2$ and $V=1.0$ models on $L_x=36,48$ cylinders with doping $\delta=11.1\%$. The same exponent $K=1.5$ is used to rescale the correlation $\Phi(r)$ for all three systems.}
\label{AFig:otherlength}
\end{figure*}

\subsection{D. Quasi-long-range CDW on six-leg cylinder}
Here we show more results about charge density properties on the six-leg cylinder. The density profile $\left< n(x,y)\right>$ obtained on the longest cylinder with $L_x=48$ is shown in Fig. \ref{AFig:cdw}(a), where we can find the density wave has spatial decaying in the bulk. The doping concentration is $\delta=11.1\%$ and interaction $V=1.0$. To precisely determine the long-distance behavior, we calculate the rung average of the density profile $\left< n(x)\right>=\frac{1}{L_y}\sum_y\left< n(x,y)\right>$. As shown in Fig. \ref{AFig:cdw}(b), the spatial decaying of the rung density $\left< n(x)\right>$ is accompanied with a complicated oscillation caused by the two-site unit cell. We further average the electron density on two sub-lattice in each unit-cell $\tilde{n}(\tilde{x})=(n_A(\tilde{x})+n_B(\tilde{x}))/2$, where $\tilde{x}$ labels the unit cell counted from the open boundary of the cylinder. The average density $\tilde{n}(\tilde{x})$ can be accurately fitted by the Friedel oscillation
\begin{eqnarray}
    \tilde{n}(\tilde{x})=A \cos(Q_c \tilde{x} +\phi) \tilde{x}^{-K_c/2} + \bar n ,
\end{eqnarray}
where $Q_c$ is the ordering momentum of CDW and $K_c$ is the Luttinger exponent characterizing the long-distance behaviors of density-density correlation function. As shown in Fig. \ref{AFig:cdw}(c), the exponent extracted from Friedel oscillation is $K_c=1.7(3)$ and the ordering momentum $Q_c$ is around $2\pi/3.5$, i.e. the wavelength of the CDW is $\sim 3.5$ unit cells. Note the exponent of SC correlation function obtained on the same system is $K_{sc}\sim 1.5 < K_{c}$, implying that the PDW ordering is the most dominant order.

\begin{figure*}[bth]
\centering
\includegraphics[width=0.8\linewidth]{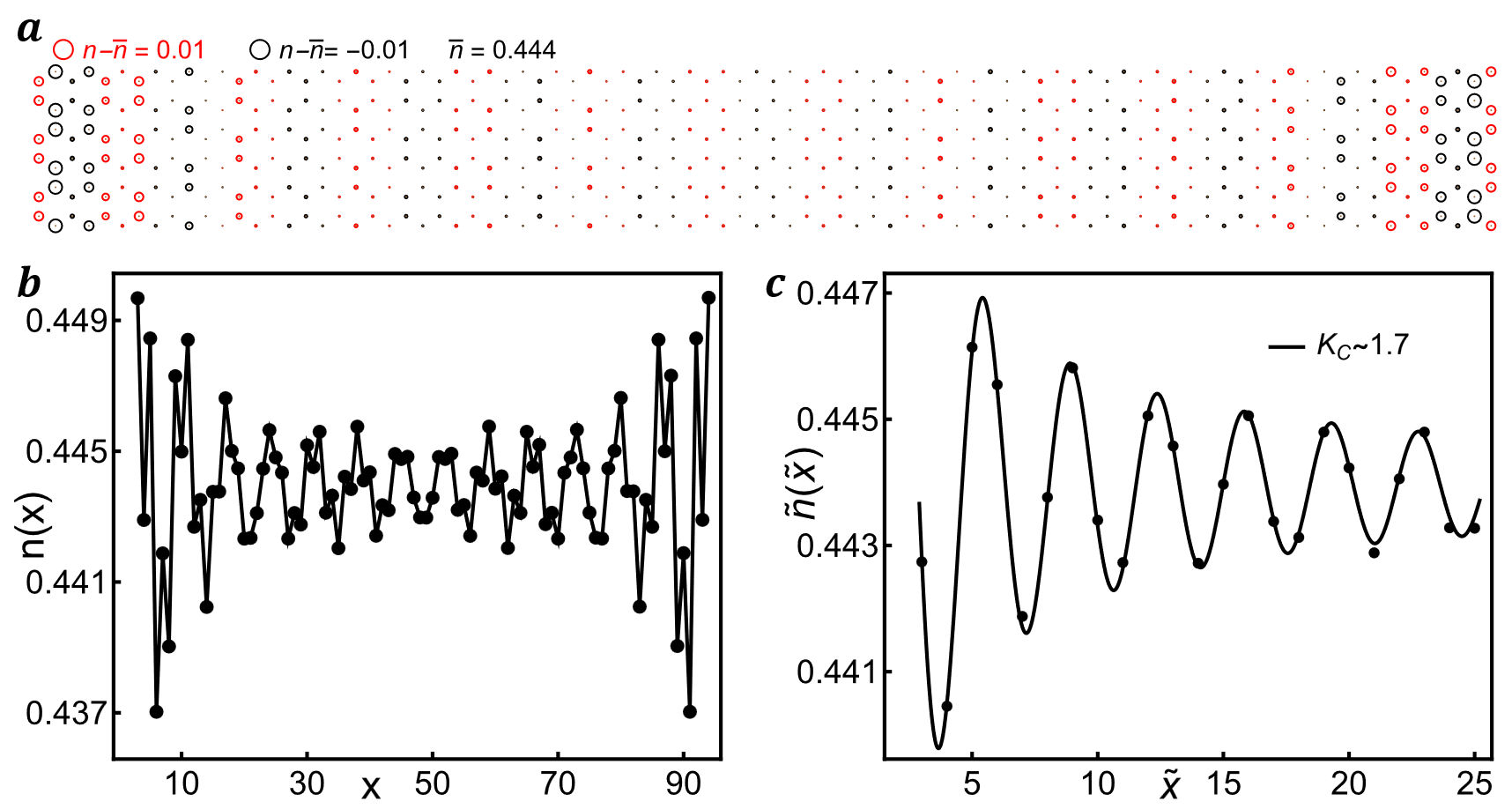}
\caption{The density properties of the $V=1.0$ model on the six-leg cylinders with $L_x=48$ and $\delta=11.1\%$: (a) the density profile of the system. Two unit cells at boundaries are omitted. (b) The rung density $n(x)$ of the system, where $x$ labels the site. (c) The average density of the unit cell $\tilde{n}(\tilde{x})$, where $\tilde{x}$ labels the unit cell. Solid line represents the fitting result of the Friedel oscillation.}
\label{AFig:cdw}
\end{figure*}

\end{widetext}

\end{document}